# Classical hybrid approaches on a transportation problem with gas emissions constraints

**Camelia-M. Pintea[1], Petrica C. Pop[1] and Mara Hajdu-Macelaru[1]**

Technical University Cluj-Napoca, North University Center
430122 Baia-Mare, Romania
cmpintea@yahoo.com, petrica.pop@ubm.ro, maram@ubm.ro

**Abstract.** In order to keep a green planet, in particular its important to limiting the pollution with gas emissions. In a specific capacitated fixed-charge transportation problem with fixed capacities for distribution centers and customers with particular demands, the objective is to keep the pollution factor in a given range while the total cost of the transportation is as low as possible. In order to solve this problem, we developed several hybrid variants of the nearest neighbor classical approach. The proposed models are analyzed on a set of instances used in the literature. The preliminary results shows that the newly approaches are attractive and appropriate for solving the described transportation problem.

## 1 Introduction

One of the new concepts in modern optimization, according to Seuring and Muller [15] is the field of sustainable supply chain design, representing a business issue affecting a company's supply chain (logistic network) in terms of environmental, risk and waste costs. The model is focused on economic and environmental aspects of greenhouse gas emissions. Santibanez-Gonzales et al. [14] described a supply chain network design problem arising in governmental agencies, where we have to decide the location of institutions as schools, hospitals, taking into account sustainable issues in the form of restrictions on the dioxide carbon equivalent emissions.

A transportation problem is a network problem [8]. Some particular transportation problems are described as fixed cost transportation problems [1, 16], that are extensions of the traditional transportation problem and are considered two kinds of costs: direct costs and fixed costs. Based on sustainability of a supply chain network [14], the current paper proposes a hybrid classic approach for solving a two-stage supply chain transportation problem. We describe the network design model and analyzed the impact of restrictions in the greenhouse gas emissions on transportation costs and in the location of facilities of the network of one layer instead of two as in [14], but involving two-stage chains as in [10,12]. The mathematical model of the problem is a mixed-integer 0-1 programming model.

One of the first multi-objective mixed-integer 0-1 model for deciding location and capacity expansion of facilities and transportation issues in a given planning horizon was developed in [9]. The profit was maximized and the environmental impact of the facility operations while satisfying the market demand for products was minimized.

The considered two-stage supply transportation problem implies a supply chain from a manufacturer who delivers items to certain distribution centers and a chain from distribution centers to a group of customers with given demands. The goal of the problem is to minimize the transportation cost, keeping the pollution, the gas emissions, as low as possible. A new constraint based on the supply chain characteristics and on the two-layers supply chain from [14] is introduced.

In order to solve complex transportation problems researchers use different heuristics [4,14] and hybrid techniques [2] . In the current paper, we develop a hybrid model using the nearest neighbor technique. In time, this approach was used to solve many difficult applications such as from data mining [6], machine learning [3], data compression [7], document retrieval [4] and statistics [5].

The current paper has five sections and its frame is organized as follows. In Sect. 2 are illustrated the two-stage supply chain fixed-charge transportation problem with the greenhouse gas emission constraint including also an example. The classical hybrid models based on Nearest Neighbor are shown in the third section followed by tests analyzes. The paper concludes with future research directions.

## 2 The two-stage supply chain fixed-charge transportation problem

The two-stage supply chain network transportation problem involves the manufacturer, some distribution centers and customers. There are several potential distribution centers (DCs) candidate for the manufacturer. There are also considered the customers whose particular demands should be satisfied by distribution centers.
The complexity of the problem involves a large number of constraints, high dimensions, uncertainties and a large number of parameters.

It is stated that the manufacturer has no capacity limitation in production. It is also assumed that each potential distribution center has distinct and different capacity in order to support the customers [10]. There are known the transportation cost from the manufacturer to a distribution center and the opening cost for each potential distribution center.

Optimizing the two-stage supply chain network transportation problem means to minimize the total cost when selecting distribution centers which supply demands of all the customers.

Fixed costs and transportation costs are parts of the total cost. For a two-chain network fixed costs are considered the opening cost for potential distribution centers and also fixed cost for transportation from distribution centers to customers.

Transportation cost implies also the two-stage costs from manufacturer to distribution centers and transportation costs from distribution centers to all customers. There are considered $m$ potential distribution centers and a number of $n$ customers, each one with a particular demand [10,12].

The following notations are used:
- $f_i$ is the opening fixed cost for distribution center $i$
- $f_{ij}$ is the fixed cost for transportation from distribution center $i$, to customer $j$
- $c_i$ is the transportation cost per unit from manufacturer to distribution center $i$
- $c_{ij}$ transportation cost per unit from distribution center $i$ to customer $j$
- $x_{ij}$ is the quantity to be transported from distribution center $i$ to customer $j$
- $a_i$ is the capacity of a distribution center $i$
- $b_j$ is the number of units demanded by customer $j$

The objective problem is to minimize the function $Z$:

$$Z = Z_{tc} + Z_{fc} \quad (1)$$

where $Z_{tc}$ is the total cost of transportation and $Z_{fc}$ is the total of the fixed costs.

$Z_{tc}$ is the total cost of transportation, including the transportation cost from manufacturer to distribution centers and from distribution centers to customers.

$$(2) \quad Z_{tc} = \sum_{m}^{i=1} c_i x_i + \sum_{m}^{i=1}\sum_{n}^{j=1} c_{ij} x_{ij}$$

where

$$(3) \quad x_i = \sum_{n}^{j=1} x_{ij}, i = 1..., m.$$

and there are two constraints:

$$(4) \quad x_{ij} \geq 0, \forall i = 1..., m, \forall j = 1..., n,$$

$$(5) \quad x_i \leq a_i, i = 1..., m.$$

$Z_{fc}$ is the sum of fixed costs, including the opening costs of distribution centers and the fixed costs from distribution centers to customers.

$$(6) \quad Z_{fc} = \sum_{m}^{i=1} f_i y_i + \sum_{m}^{i=1}\sum_{n}^{j=1} f_{ij} y_{ij}$$

where

$$(7) \quad y_i = \begin{cases} 1, \sum_{n}^{j=1} x_{ij} \geq 0 \\ 0, \sum_{n}^{j=1} x_{ij} = 0 \end{cases} \forall i = 1..., m,$$

$$(8) \quad y_{ij} = \begin{cases} 1, x_{ij} \geq 0 \\ 0, x_{ij} = 0 \end{cases} \forall i = 1..., m, \forall j = 1..., n.$$

The new constraint of the problem based on [14] follows:

$$(9) \quad \sum_{m}^{i=1} \alpha' a_i x_i + \sum_{m}^{i=1} \beta'_i a_i c_i x_i + \sum_{m}^{i=1}\sum_{n}^{j=1} \alpha_i b_j x_{ij} + \sum_{m}^{i=1}\sum_{n}^{j=1} \beta_{ij} b_j c_{ij} x_{ij} \leq GHG,$$

where $\alpha_i$ = GHG emissions factor of a facility located at distribution center, in tons of $CO_2e$ per unit demand, $\alpha'$ = GHG emissions factor of a facility located at manufacturer, in tons of $CO_2e$ per unit demand, $\beta_{ij}$ = GHG emissions factor per unit distance and per unit demand between distribution center $i$ and customer location $j$, in tons of $CO_2e$ per km and unit demand and $\beta'_i$ = GHG emissions factor per unit distance and per unit demand between manufacturer and distribution center $i$, in tons of $CO_2e$ per km and unit demand.

The novelty of our paper is that we consider as well the chain from manufacturer to distribution center, that is why the parameters $\alpha'$ and $\beta'_i$ are also involved in the (9) inequality of sum emission factor.

Fig. 1 shows an example of a supply chain network as in [10].

Using the example from [10] are considered three potential distribution centers and five costumers. Data's properties are detailed in [10]. The best solution for this example is choosing distribution centers *2* and *3* as follows: the customers *1* and *4* are served from distribution center *2* and customers *2, 3* and *5* receive their demands from distribution center *3*.

The results of computing the total cost for this allocation is *3625*.

For the current paper purpose is investigated the *GHG* value of the greenhouse gas, *GHG*, emissions produced by the transportation and the operation of the facilities. For this particular problem all $\alpha$ values are considered equal and all $\beta$ values are equal too.

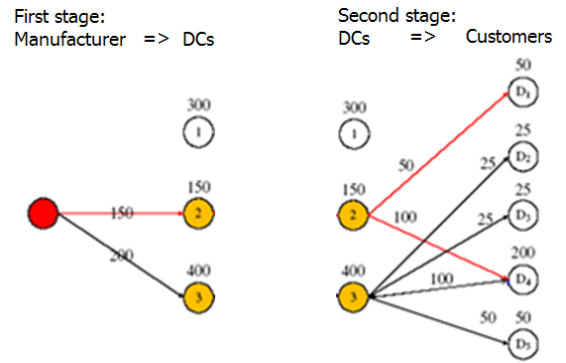

**Fig. 1** An example for a two stage supply chain network [10]

The results of computing the total GHG for the given solution:

The gas emissions from the manufacturer to DCs is
$\beta \cdot (3 \cdot 150 + 3 \cdot 200) + \alpha \cdot (150 + 200) = \beta \cdot 1050 + \alpha \cdot 350$

The gas emissions from DCs to customers is
$\beta \cdot (3 \cdot 50 + 1 \cdot 100 + 2 \cdot 25 + 5 \cdot 25 + 3 \cdot 100 + 4 \cdot 50) + \alpha \cdot (50 + 100 + 25 + 25 + 100 + 50) = \beta \cdot 925 + \alpha \cdot 350$

The total emission is $\beta \cdot 1975 + \alpha \cdot 700$. Considering $\alpha = 1$ and $\beta = 2$, the total emission is *4650*. Considering $\alpha = 0.01$ and $\beta = 0.02$, the total GHG emission is *46.5*.

In order to get closer to a real approach to a sustainable supply chain design with the action to reduce and to control the pollution, the total emissions is limited by a given value denoted GHG. In this particular case let consider the following cases:
- if *GHG=50* then *46,5<50* and therefore the greenhouse gas emissions are in the limits
- if *GHG=40* then *46,5>40* and therefore the pollution is out of the limits and some restrictions should be taken by authorities.

## 3 Classical hybrid models based on *Nearest Neighbor*

In order to reduce the greenhouse gas *(GHG)* emissions produced by the transportation and the operation of the facilities for a two-stage supply chain transportation problem it is involved *Nearest Neighbor* technique. In [12] are introduced several hybrid heuristic approaches based on *Nearest Neighbor* technique.

For the fixed-charged transportation problem, with two stages of a supply chain network, *Nearest Neighbor* algorithm could be applied when are chosen the potential distribution center and also when are chosen the best edges from distribution centers to customers. There are used two variants for this supply chain. DX-randomly chosen the potential distribution centers and DY-with the best probability based on the total request and distribution centers capacities as in [12]:

(10) $p_i = \dfrac{a_i}{request}$.

(11) $p_i = \dfrac{a_i}{xcont_i \cdot request}$.

(12) $p_i = \dfrac{xcont_i}{a_i}$.

where $xcont_i$ is the number of nonzero quantities to be transported from distribution center *i* to customer *j*, when $x_{ij}>0$.

## 4 Tests and results

The algorithms described in Sect. 3, used to solve the two-stage supply chain fixed-charged transportation problem were coded in java and compiled under Linux. For testing our implementation were generated *9 instances, 3 instances for each dimension considered (10 x 10, 10 x 30, 30 x 100)* as in [12]. For every instance a two-stage supply chain with one layer is considered. Hybrid considered algorithm are choosing in different ways the distribution center starting from the manufacturer.

*HNN-DX* is the algorithm where a distribution center is chooses randomly and *HNN-DY 10-12* are algorithms with distribution center chosen with a probability specified in equations (10), (11) and (12).

The parameters used in Sect. 2, inequality (9) are: the total emissions limit *GHG* considered *150,000* and the greenhouse gas emissions coefficients *α'=α=0.02* and *β'=β=0.04*. Table 1 show that the optimal objective of greenhouse gas emissions values was obtained by *HNN-DY10 based on* an unpaired t-test of other considered algorithms versus *HNN-DY10*.

**Table 1** Results of unpaired t-test results, based on HNN-DY10 results

|  | NN. | HNN-DY11 | HNN-DY12 | HNN-DX |
|---|---|---|---|---|
| t-test result t the probability of the | 0.348 | 0.358 | 0.64 | 0.875 |
| result, assuming the null hypothesis | 0.73 | 0.72 | 0.475 | 0.39 |

The difference is considered to be not statistically significant, based on the unpaired t-test results, the value closer to *HNN-DY10* are in order: *NN, HNN-DY11, HNN-DY12* and *HNN-DX*. An important observation is that not always the cost of the transportation is proportional with greenhouse gas emissions.

Further work should be done considering other hybrid heuristics as bio-inspired ones [2,11,13] in order to give a better support for transportation problems. Another perspective of the two-stage supply chain fixed-charged problem should be considered when using multi-layer facilities.

## 5 Conclusions and future work

A two-stage supply chain network design problem that arises in the public sector considering sustainable constraints in the form of restrictions on the dioxide carbon equivalent emissions is considered. Hybrid heuristics are involved, using different probabilities in order to find the suitable distribution center starting from a given manufacturer. The entire request from all customers should be satisfied. We observed that the cost of the transportation is not always proportional to the greenhouse gas emissions.